# Round-The-Clock Urine Sugar Monitoring System for Diabetic Patients

Pramit Ghosh
Department of Computer Science and Engineering,
RCC Institute of Information Technology,
Kolkata 700015, India
pramitghosh2002@yahoo.co.in

Debotosh Bhattacharjee, Mita Nasipuri, and Dipak Kumar Basu
Department of Computer Science and Engineering,
Jadavpur University,
Kolkata 700032, India
debotoshb@hotmail.com, mita_nasipuri@gmail.com, dipakkbasu@gmail.com

*Abstract*— **It is known that diabetes can not be cured completely, but it can be controlled. The objective of this work is to provide an automatic system that will be able to help the diabetic patient to control the blood sugar. This system measures the blood sugar level of the people from their urine round-the-clock. A recorded message based on this input may be displayed so that apart from patient himself others can be informed about his/her present sugar level. That should help him/her in taking medicine; controlling diet etc. This work is an application of image processing and fuzzy logic. It is known that Benedict's reagent changes its colour based on the sugar level. This colour change information is sensed by the transducer and fed to the fuzzy logic unit for decision making.**

## I. INTRODUCTION

Diabetes is a condition in which the pancreas no longer produces enough insulin or when cells stop responding to the insulin that is produced, so that glucose in the blood cannot be absorbed into the cells of the body. Symptoms include frequent urination, tiredness, excessive thirst, and hunger. The treatment includes changes in diet, oral medications, and in some cases, daily injections of insulin. [1].

The total number of diabetes patients in the world is expected to grow from 246 Million in 2006 to 380 Million by 2025[2]. Diabetes experts believe that blood-sugar level above 240 mg/dl is unacceptable and dangerous; however, the ideal level is 80 to 120 mg/dl range. Uncontrolled diabetes can give rise to many complications, which are either acute or short term; or chronic or long term like Cataracts, Diabetic Retinopathy, diabetic nephropathy etc [3].

This work is basically a round the clock blood sugar monitoring system for a diabetic patient. This work is an application of image processing and fuzzy logic. It is an alternative approach of blood sugar level testing from the urine rather than from the blood [4] because people are not interested to test their blood several times in the day, until they are feeling bad because blood sample collection process is painful.

The rest of the paper is arranged as follows: section II contains the detail design including hardware design and algorithms; section III focuses the results and performance of the system and section IV is used to discuss the future scope and overall performance of the system.

## II. SYSTEM DETAILS

The entire work is an amalgamation of chemistry, mechanical and control engineering, image processing and fuzzy logic. The next sections will explain each of these.

### A. Benedict's Reagent

Benedict's reagent [5] changes its colour based on the sugar level. Blue colour denotes that urine sugar is Nil, Green means approximate blood sugar level is 0.1- 0.5 g/dl denoted by +, in Yellow sugar level 0.5- 1.0 g/dl marked as ++, Orange colour marked as +++ and sugar level 1.0- 1.5 g/dl, Brick red colour means sugar is very high 1.5- 2.0 g/dl marked as ++++ [6] however urine sugar level depends on the renal threshold. Figure 1 shows the colour.

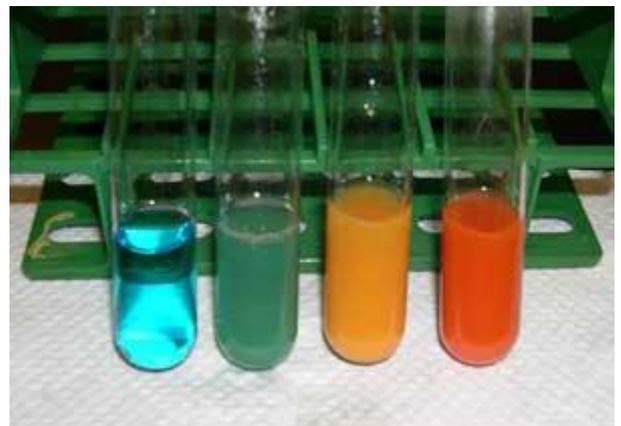

Figure 1. Different colours of Benedict's reagent

### B. The Architecture of the System

The block diagram of Figure 2 shows a small quantity of urine is to be collected from toilet outlet pipe (A) with 3ml of

Benedict reagent (B) is to be mixed. After that temperature is applied through a heater (H) up to the boiling point. After boiling wait for a few minutes for cooling and check colour of the solution. A colour sensor is mounted to record colour and produce output accordingly. Output thus found is given as input to a processing module to estimate level of sugar.

In Figure 2 V1, V2 and V3 are three computer controlled valve, namely solenoid valve [7]. Computer can not control the solenoid valve directly. To control that type of valve driver circuit is required. The computer output from parallel port is denoted by p0.1 and it is fed to the optocoupler (MCT2E) for electrical isolation. However these parallel port will be replaced by USB and micro controller combination in the next version. The output of the optocoupler is fed to the relay via a current amplifier, because the current provided by the MCT2E is not sufficient to drive relay. When the relay coil is switched off then a high voltage is generated by the relay coil. To protect the circuit, a diode (D1) is used as is shown in Figure 3. V1 opens to collect urine from toilet outlet (A). V2 allows Benedict reagent to enter in the chamber. H is the heater to boil the mixture. It is also controlled by the computer. V3 opens to drain, the liquid after sensing the colour. The entire process is being discussed with the help of a flow diagram given in Fig 4.

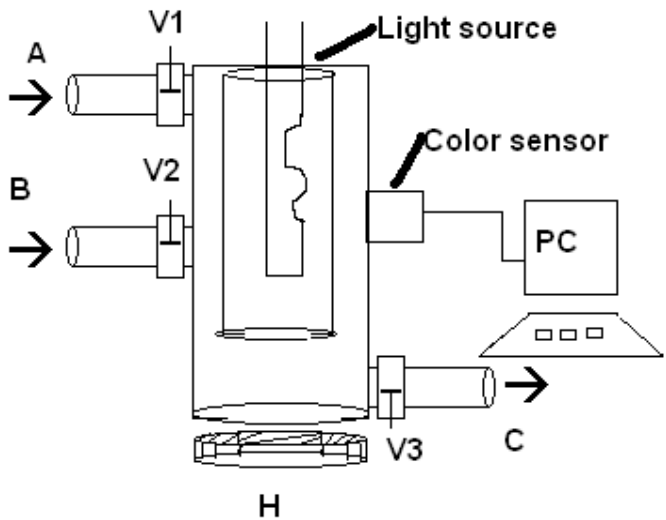

Figure 2. Block diagram of the system.

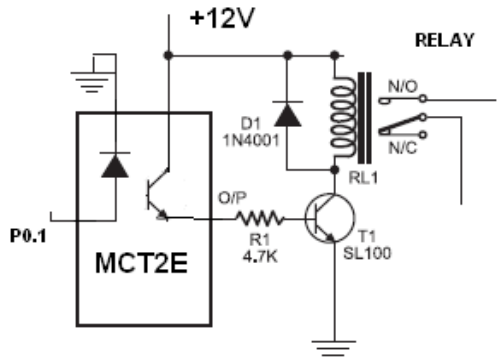

Figure 3. Circuit diagram of the driver.

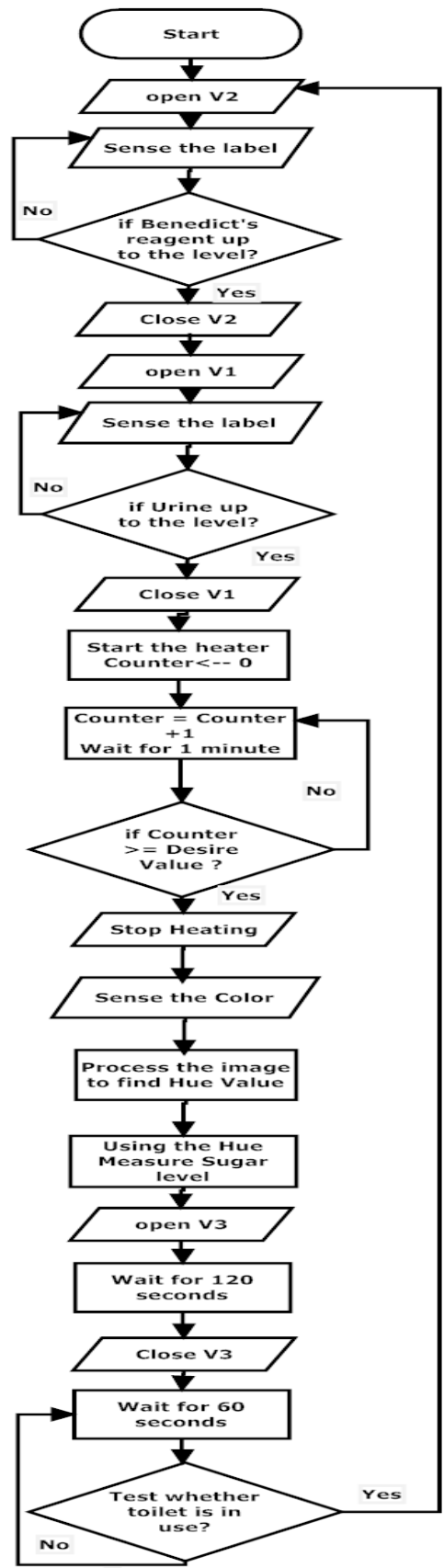

Figure 4. Flow diagram of the work

## C. Colour Sensing

The colour change of the Benedict's reagent is uniform but the image captured by the transducer has multiple similar colours that must be treated as noise. To reduce this type of noise, convolution based filtering [8] is used separately on the Red, Green and blue components of the sensed image using equation 1.

$$g(x, y) = \sum_{s=-a}^{a} \sum_{t=-b}^{b} w(s,t) f(x+s, y+t) \quad (1)$$

where w is the convolution mask, in which all co-efficient values are 1/(m×n), of size m×n. m=2a+1 and n=2b+1, where a and b are nonnegative integers. The co-efficient w(0,0) coincides with image value of f in the (x,y) position, indicating that the mask is centered at (x,y) when the computation of the sum of products takes place. g(x,y) contains the response of the filter. x and y represent the row and column indexes of the image. After applying the filter the red, green and blue components are assembled again to build actual image. The effect of applying filter is shown in Figure 5.

The next step is to extract colour information. But the problem is that RGB component contains not only the colour information but also the colour intensity i.e. the RGB values are different for light blue, dark blue and navy blue. So from RGB components it is very difficult to identify the colour. To overcome this problem HSI [9] colour format is used. Where H stands for Hue (Pure colour), S for saturation, i.e. pure colour is diluted by white light and I for intensity (Gray level)

The RGB to HSI colour space conversion process is performed using the equations.

$$H = \begin{cases} \theta \dots\dots\dots if\ ..B \leq G \\ 360 - \theta\ ..otherwise \end{cases} \quad (2)$$

$$\theta = \cos^{-1}\left\{\frac{\frac{1}{2}[(R-G)+(R-B)]}{\sqrt{[(R-G)(R-G)+(R-B)(G-B)]}}\right\} \quad (3)$$

$$S = 1 - \frac{3}{(R+G+B)}[\min(R,G,B)] \quad (4)$$

$$I = \frac{1}{3}(R+G+B) \quad (5)$$

R,G,B are the Red, Green and Blue components of a pixel.

Algorithm for Hue value collection from the image.

RGB stands for Red, Green and Blue components of the image. H is a matrix that contains the Hue values of the image. "Row" is a vector. minimum () is a function which returns the minimum value from its arguments.

Step 1: Apply equation 3 on each pixel of the RGB image and store the value in H matrix.

Step 2: Sort the entire row's elements of the H matrix and find the middle element and store it in the vector named "Row".

Step 3: Sort the "Row" vector and find the middle value and store it in "Hue" variable.

Step 4: As it is known that the value of Hue varies from 0 to 1 so

ActualHue = minimum (Hue, 1- Hue).

Step 5: Return the ActualHue. Stop.

## D. Decision Making

It is very difficult to define crisp set to identify the pure colour from Hue value. To solve this problem fuzzy set is used. In this system the fuzzy membership [10] function is build from a reference data set. The advantage of this membership function is flexibility, i.e. by changing the reference data set or training set the response of the membership function (equation 6) can easily be changed.

$$f(X) = \begin{cases} 0 \dots\dots\ if\ .X < .\min(DataSet)\ or \\ \dots\dots\dots X > \max(DataSet) \\ \sum_{i=1}^{n} Yi . \prod_{j=1, j\neq i}^{n} \frac{(X-Xi)}{(Xi-Xj)} \dots otherwise \end{cases} \quad (6)$$

where f(X) is the response of the membership function, X is the input value, DataSet is the set of training set. n is the number of data in the DataSet, Xi is the i th member of the DataSet. Xi stands for input and Yi for corresponding output. Similarly Xj denotes the j th input data set.

In urine glucose test using Benedict's reagent pathological interest is on blue, green, yellow and orange colour. So only this four colour's test set is required, to build separate membership functions. Using majority voting technique, upon the outcome of this membership functions, the final decision is to be taken.

The next section describes the Algorithm to find the fuzzy membership function: HueVec is a vector that contains the hue values of a test set, val is an another vector that contains the response of the Hue value. The dimension of the HueVec vector and val vector must be same. X is an input hue value whose colour has to be detected. colarValue is a variable which contains the percentage of the presence of that colour.

Step 1: If X is less than minimum of HueVec then

   colarValue = 0. Exit.

Step 2: If X is greater than maximum of HueVec then colarValue = 0. Exit.

Step 3: Repeat step 4 to step 7 until the number of elements in HueVec

Step 4: Set product=1;

Step 5: Repeat step 6 until the number of elements in HueVec

Step 6: If the index in the step 3 and index in the step 5 are not equal then product = product *((X - HueVec (index in the step 5))/( HueVec (index in the step 3) – HueVec( index in the step 5)));

Step 7: Set colarValue = colarValue + product* val (index in the step 3);

Step 8: Stop end.

As it is round-the-clock monitoring system so a question arise how to distinguish patient from other family members. To solve the problem a philosophy is used. Generally, in case of nuclear family i.e., husband, wife and children it is possible to distinguish each of the family member from others by his/her weight. A single element of piezoelectric co-polymer film weighing machine is to be placed just at the entry point of the toilet room, without introducing any disturbance in smooth walking. Once the weights of all the individuals (which are distinct) are recorded, computer can easily understand through the output of the weighing machine about the person presently using the toilet.

### III. SIMULATION AND RESULT

For simulation MatLab [11] is used. Figure 5 shows the sensed image along with the noise removed image.

Figure 5. The sensed image along with the noise removed image.

After that hue (pure colour) is extracted from the filtered image using equation 2 and value is 0.0257 shown in figure 6

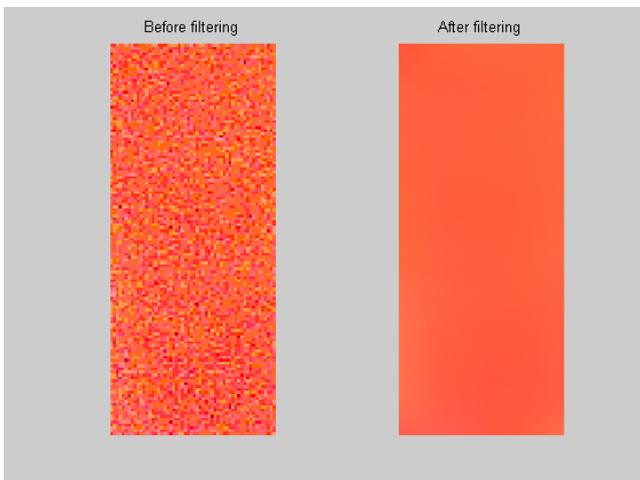

Figure 6. The hue value of figure 5

Table 1 shows the training sets used by the equation 6 to build the red, yellow, green and blue fuzzy membership functions. In the table 1 there are 8 columns, the $1^{st}$ one shows the hue value and the $2^{nd}$ column shows the percentage of the presence of red colour for that hue value, the $1^{st}$ and $2^{nd}$ column are used to build the fuzzy member function for red colour using equation 6; similarly $3^{rd}$ and $4^{th}$ for yellow, $5^{th}$ and $6^{th}$ for green and finally $7^{th}$ and $8^{th}$ column for blue component.

TABLE I
TREANING SET

| Training sets for Red | | Training sets for Yellow | | Training sets for Green | | Training sets for Blue | |
|---|---|---|---|---|---|---|---|
| Hue | % | Hue | % | Hue | % | Hue | % |
| 0.007 | 100 | 0.039 | 20 | 0.184 | 10 | 0.417 | 20 |
| 0.039 | 80 | 0.060 | 30 | 0.192 | 20 | 0.452 | 30 |
| 0.060 | 70 | 0.082 | 40 | 0.214 | 70 | 0.481 | 60 |
| 0.082 | 60 | 0.106 | 50 | 0.245 | 80 | 0.5 | 80 |
| 0.106 | 50 | 0.127 | 60 | 0.272 | 100 | 0.510 | 100 |
| 0.127 | 40 | 0.148 | 70 | 0.314 | 100 | 0.534 | 100 |
| 0.148 | 30 | 0.166 | 90 | 0.333 | 100 | 0 | 0 |
| 0.166 | 10 | 0.184 | 90 | 0.394 | 90 | 0 | 0 |
| 0 | 0 | 0.192 | 80 | 0.417 | 80 | 0 | 0 |
| 0 | 0 | 0.214 | 30 | 0.452 | 70 | 0 | 0 |
| 0 | 0 | 0.245 | 20 | 0.481 | 40 | 0 | 0 |
| 0 | 0 | 0 | 0 | 0.5 | 20 | 0 | 0 |

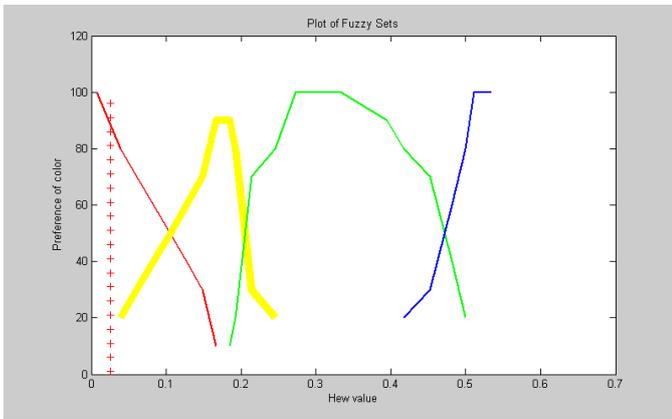

Figure 7. The membership function. Along with detected value.

In Figure 7 the red vertical line shows the position of the hue value of the colour of the test sample, shown in Figure 5. The red line shows the membership function's response decaying against hue value. Yellow line shows the response of the yellow fuzzy set and similarly green and blue lines for the response of green and blue fuzzy set. Figure 8 shows the numerical values, in which red component has the highest value 85.6541. So the final decision is that the colour is Red. According to the pathologist the approximate blood sugar level will be 290 -350 mg/dl for that particular colour in Figure 5.

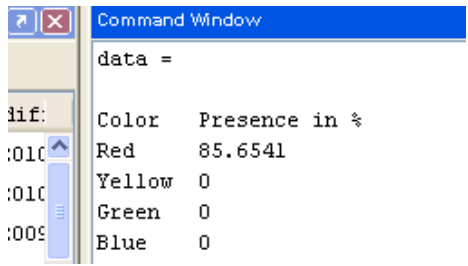

Figure 8. The values from membership functions

For testing purpose 130 samples are used in which 126 give satisfactory results. So the error rate is 3.07%

## IV. CONCLUSION

This system is efficient to detect blood sugar level and also cheap. It is easy to install. It can reduce the probability of glaucoma, kidney problem etc. by early detection of high blood sugar level. However this approach provides 96.93% accuracy, but the algorithms for decision making should be changed to increase the accuracy. Future plan of this work is to enhance the learning as well as decision making algorithms. If more training set is used then the decision will be more accurate. As a conclusion it can be said that this system has a good market value if it is advertised properly.

ACKNOWLEDGMENT

Authors are thankful to Dr. Abhijit Sen for providing pathological data and the "Center for Microprocessor Application for Training Education and Research", of Computer Science & Engineering Department, Jadavpur University, for providing infrastructural facilities during progress of the work. Dr. Dipak Kumar Basu acknowledges the thanks to AICTE, New Delhi for providing an Emeritus fellowship.